\documentclass[prl,reprint,showpacs]{revtex4-1}

\usepackage{epsfig}
\usepackage{graphics}
\usepackage{amsmath}
\usepackage{color}

\begin{document}

\author{Yoav Sagi, Miri Brook, Ido Almog and Nir Davidson}
\affiliation{Department of Physics of Complex Systems, Weizmann Institute of Science, Rehovot 76100, Israel}

\title{Observation of anomalous diffusion and fractional self-similarity in one dimension}

\begin{abstract}
We experimentally study anomalous diffusion of ultra-cold atoms in a one dimensional polarization optical lattice. The atomic spatial distribution is recorded at different times and its dynamics and shape is analyzed. We find that the width of the cloud exhibits a power-law time dependence with an exponent that depends on the lattice depth. Moreover, the distribution exhibits fractional self-similarity with the same characteristic exponent. The self-similar shape of the distribution is found to be well-fitted by a L\'{e}vy distribution, but with a characteristic exponent that differs from the temporal one. Numerical simulations suggest that this is due to long trapping times in the lattice and correlations between the atom's velocity and flight duration.
\end{abstract}
\maketitle

Diffusion is a phenomenon encountered in almost every branch of physics. Its ubiquitousness stems from the central limit theorem, which states that a sum of random variables is distributed normally as the number of addends increases. It holds, however, only when the distribution of the variables has a finite variance. When this assumption does not hold, i.e. for heavy-tailed distributions with asymptotic power-law behavior with an exponent $\alpha+1$ with $0<\alpha<2$, the sum converges instead to a L\'{e}vy distribution $L_{\alpha}$. A diffusion process which results in such non-Gaussian spatial distribution is usually regarded as anomalous \cite{Bouchaud1990127}. Heavy-tailed distributions are found in many fields, from animals foraging strategies \cite{Humphries2010}, to the prices of commodities and stocks \cite{Mandelbrot_1963}. In physics they emerge in situations where there is no characteristic length scale, such as near phase transitions \cite{JPSJ.66.314}, in turbulent flow \cite{PhysRevLett.71.3975}, in quantum phase diffusion \cite{PhysRevLett.54.616} or when a system is out of thermal equilibrium \cite{levy_statistics_and_laser_cooling_book}. In fact, anomalous transport properties are intimately linked to non-linear chaotic dynamics which naturally appears in many physical systems \cite{klafter_nature_1993,PhysRevA.75.063428}.

A simple diffusion model we have in mind is that of particles in real space, each having a velocity which fluctuates in time due to interaction with a bath. After some time the particles' position is distributed as $W(x,t)$. The characteristic width of the ensemble, e.g. the full width at half the maximum (FWHM), usually scales as a power-law $t^{1/\alpha}$. Anomalous diffusion can arise when the distribution of velocities has heavy-tails, and almost always results in $\alpha\neq2$ \cite{Bouchaud1990127}. The theoretical challenge is to connect the microscopic physics to the evolution of the distribution $W(x,t)$. The experimental challenge is to measure these distributions in a well controlled and isolated environment. In this work we meet this challenge by measuring anomalous diffusion of laser cooled atoms in a polarization optical lattice \cite{DALIBARD1989,limits_sisyphus_1991,Lutz2003,Marksteiner1996}. In this system the steady state atomic velocity distribution was shown both theoretically and experimentally to follow a power-law, with an exponent that depends on the lattice depth \cite{limits_sisyphus_1991,Lutz2003,PhysRevLett.96.110601}. It was also predicted that the real space diffusive motion of the atoms in such a lattice is anomalous for a wide range of lattice parameters \cite{Marksteiner1996}. The onset of anomalous transport characteristics was observed with a single trapped ion \cite{Katori1997}. However, a measurement of the distribution $W(x,t)$ and its anomalous dynamics was not reported to date.

\begin{figure}
    \begin{center}
    \includegraphics[width=8cm]{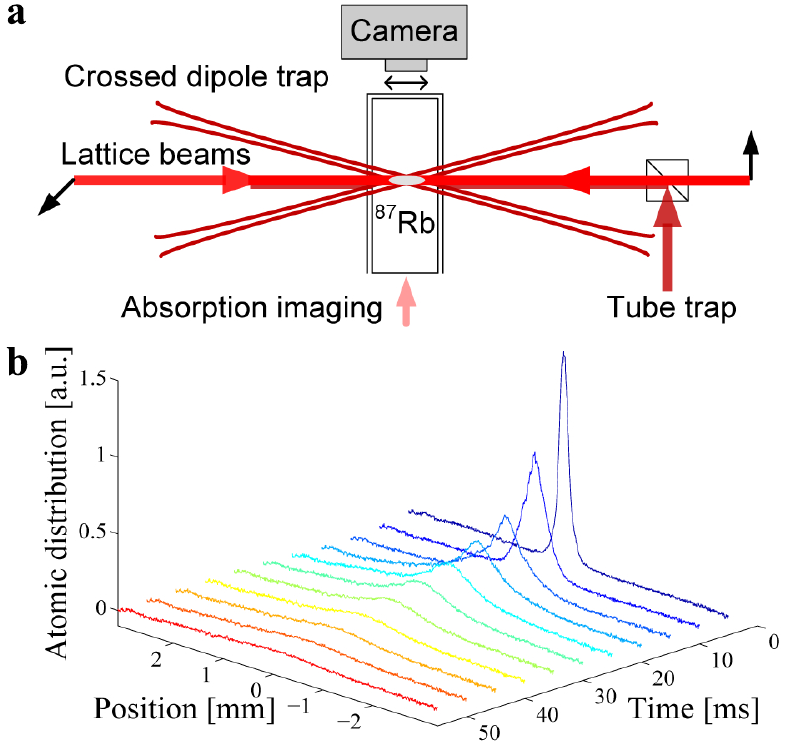}
    \end{center}\caption{\textbf{a}, The experimental setup as seen from above. The atoms are released from a small crossed dipole trap into a second far-off-resonance trap with a very elongated aspect ratio which confine their motion to a single direction (tube trap). This trap is superimposed with the two counter-propagating Sisyphus lattice beams which induce the diffusive motion. After a chosen time, the atomic density is recorded using absorption imaging technique. Each image gives the column two dimensional density, and we integrate over the radial directions to obtain the axial atomic distribution. \textbf{b}, Atomic distributions after different diffusion times in a $4.8 E_r$ deep lattice. Each measurement is repeated 20 times and averaged to improve the signal to noise ratio.}\label{figure1}
\end{figure}

Here we report such a measurement in one dimension with an ensemble of ultra-cold $^{87}$Rb atoms. By setting out with a very small atomic cloud and recording the longitudinal density distribution after different waiting times, we are able to directly measure anomalous dynamics in the lattice. We find that the width of the distribution exhibits a power law time dependence, from which a characteristic exponent can be extracted. The value of this exponent depends on the lattice depth. Furthermore, we show that the density distribution at different times exhibits self-similarity with the same characteristic exponent. The self-similar shape of the distribution is found to be very well fitted by a L\'{e}vy function. However, the characteristic exponent extracted from this fit is significantly smaller than the exponent extracted from the dynamics. We investigate this point using classical and quantum Monte-Carlo simulations and find that it originates from correlations between the atom's velocity and the corresponding time it spends un-trapped in a particular lattice site, combined with heavy-tailed distribution of the durations in which it is trapped by the lattice.

The apparatus is depicted schematically in figure \ref{figure1}a. In each experiment, $\sim 1.5\cdot10^6$ $^{87}Rb$ atoms are prepared in a cigar shaped crossed dipole trap with an aspect ratio of 1:3.9 and a radial oscillation frequency of $2\pi\cdot 420$Hz (for more details regarding the apparatus see Ref. \cite{PhysRevLett.104.253003}). The Sisyphus lattice is created by two counter-propagating lattice beams with identical wavelength and orthogonal linear polarizations \cite{DALIBARD1989}. These beams originate from an external cavity diode laser whose frequency is locked to an atomic transition and detuned $-66$MHz relative to the transition between states $5^2S_{1/2}, F=2$ and $5^2P_{3/2}, F'=3$. The atomic cloud has its long axis aligned parallel to the lattice beams. The lattice depth is calculated from the measured beam's intensity and waist of $1.1mm$, and using a saturation intensity of $3.6mW/cm^2$. Before switching on the lattice, the atomic cloud has a temperature of $12\mu K$ and a maximum phase space density of $1.7\cdot 10^{-3}$. To counteract gravitation and improve the signal to noise ratio, the motion of the atoms is confined to the longitudinal direction by superimposing the lattice with a ``tube'' trap, namely a Gaussian beam with a waist of $\sim 120\mu m$ such that the Rayleigh range is much larger than typical diffusion distances we record. Both the crossed dipole trap and tube trap originate from a single frequency Ytterbium fiber laser at a wavelength of $1.06\mu m$, but their frequencies are shifted relative to each other by more than $20$MHz to prevent standing waves. The power of the tube trap beam is $10W$.

The lattice is turned on $1ms$ before the crossed dipole trap is turned off, during which the atoms equilibrate with the lattice. The average rate of photons scattered by each atom from the lattice is $\sim 10^4 s^{-1}$, much larger than the initial maximal average elastic collisions rate of $190 s^{-1}$. This is important since elastic collisions lead to an undesirable thermal equilibrium. We define $t=0$ as the time at which the crossed dipole trap is switched off and the atoms start diffusing in the lattice. The initial size of the cloud is $\sim 200\mu m$, much smaller than the typical diffusion distances. We take a series of absorption images after different waiting times, an example of which in a $4.8 E_r$ deep lattice is depicted in Figure \ref{figure1}b.

A convenient and useful theoretical framework which describes a broad range of anomalous diffusion processes is the fractional diffusion equation (FDE) \cite{Metzler20001}:
\begin{equation}\label{FDE}
\frac{\partial W(x,t)}{\partial t}=D_t^{1-\beta} K_\beta^\mu D_x^\mu W(x,t) \ \ ,
\end{equation}
where $W(x,t)$ is the atomic distribution at time $t$, $D_x^\mu$ is the Weyl operator describing a fractional derivative in space and similarly $D_t^{1-\beta}$ is the fractional time derivative. $K_\beta^\mu$ is a generalized diffusion constant having the dimensions of $cm^\mu/s^{\beta}$. For $\mu=2$ and $\beta=1$ this equation reduces to a normal diffusion equation. $\mu<2$ corresponds to long spatial jumps (also referred to as L\'{e}vy flights), whereas $\beta<1$ corresponds to long dwelling times between jumps. The solution for the kernel $G_{\beta \mu}(x,t)$ of this equation can be written in terms of Fox functions \cite{Mainardi2005}. A general property of the kernel is its time and space scaling \cite{Metzler20001,Mainardi2005}:
\begin{equation}\label{scaling}
G_{\beta \mu}(x,t)=t^{-\beta/\mu}\tilde{G}_{\beta \mu}(x t^{-\beta/\mu}) \ \ ,
\end{equation}
where $\tilde{G}_{\beta \mu}$ is the reduced kernel function.

\begin{figure}
    \begin{center}
    \includegraphics[width=8cm]{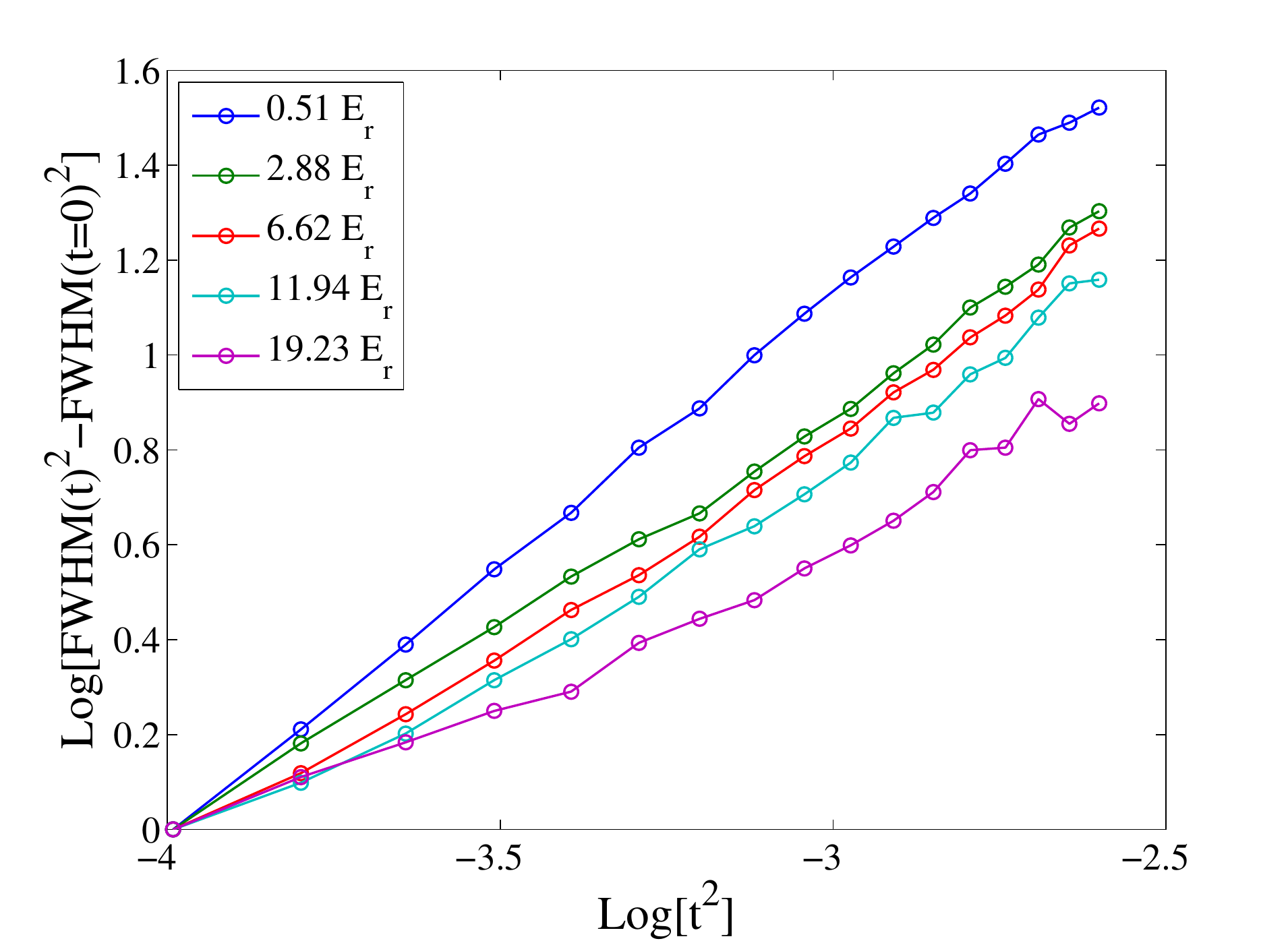}
    \end{center}\caption{A log-log graph of the FWHM squared as a function of the time squared, for different lattice depths. To facilitate the comparison of the slopes, all graphs were shifted in the y-direction to cross at a single point. The linear behavior establishes the power-law time scaling of the width of the atomic distribution. The FWHM is found by fitting the atomic distributions to a general function from which we extract the width. In very shallow lattices (below $\sim 1.7E_r$) we notice an excess density in the center of the distribution which is probably due to corrugation in the tube trap potential. To avoid systematic errors due to this effect, we exclude in this range of lattice depths the central part of the distribution when fitting and extracting the FWHM.}\label{log_log_of_fwhm}
\end{figure}

\begin{figure}
    \begin{center}
    \includegraphics[width=7.5cm]{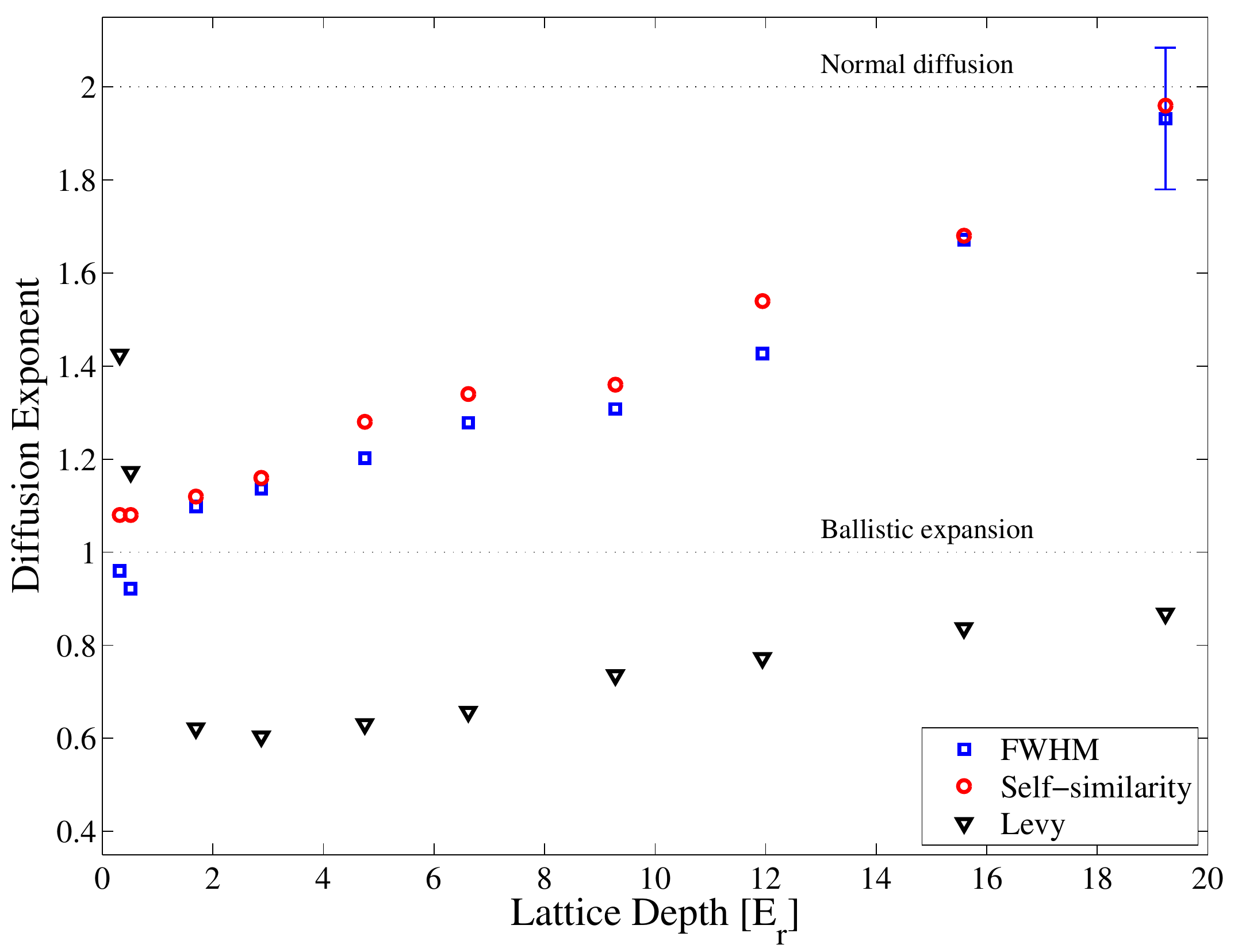}
    \end{center}\caption{The diffusion exponent $\alpha$ as a function of the lattice depth. The exponent is extracted by three different methods; The (blue) squares are obtained by fitting the FWHM scaling to a power-law $FWHM\sim t^{1/\alpha}$ (the errorbar represents a $95\%$ confidence level). The (red) circles are obtained from the measure of the self-similar transformation. The (black) triangles are found by fitting the distributions taken after $30ms$ with a L\'{e}vy $L_\alpha$ function.}\label{diffusion_exponent}
\end{figure}

One conclusion that can be immediately drawn from Eq. \ref{scaling} is that the typical width of the distribution should scale as $t^{1/\alpha}$, with a dynamical diffusion exponent given by $\alpha=\mu/\beta$. To test this, we plot in figure \ref{log_log_of_fwhm} on a log-log scale the FWHM extracted from the data as a function of time for different lattice depths. The curves are approximately linear, indicating that indeed the width scales as a power-law in time. Furthermore, the slope of each curve is different, showing that the diffusion exponent depends on the lattice depth, as predicted \cite{limits_sisyphus_1991,Lutz2003}. We fit each of these curves with a line, and from its slope we extract the diffusion exponent. The results are plotted as squares in figure \ref{diffusion_exponent}, and demonstrate that the whole range of fractional diffusion exponents is accessible in our experiment by changing the lattice beams power. Note that since in the radial direction there is no cooling, atoms are eventually lost from the trap due to spontaneous emission. Owing to the large Rayleigh range of the tube trap, there is almost no mixing between the axial and radial velocity distributions and therefor and therefore the radial loss does not distort the axial spatial distribution. Nevertheless, we use the data only as long as at least $10\%$ of the atoms remain in the trap. Increasing this number to $30\%$ changes only the 3 largest points in figure \ref{diffusion_exponent} by up to $25\%$.

\begin{figure}
    \begin{center}
    \includegraphics[width=7.5cm]{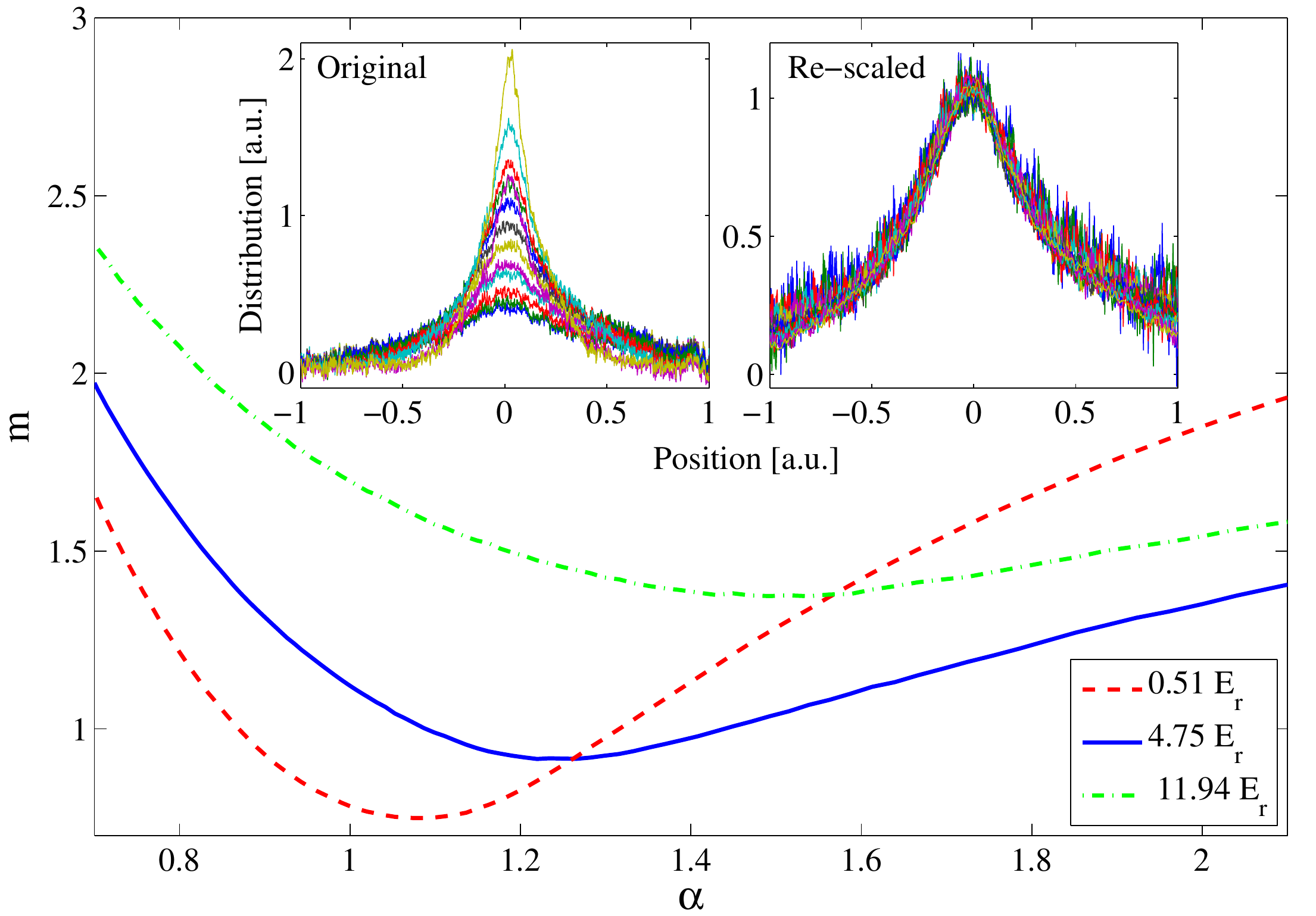}
    \end{center}\caption{The measure of the self-similarity, $m$, as a function of the diffusion exponent, $\alpha$, for three different lattice depths. Each point is calculated from 13 distributions taken after $10-40ms$ of diffusion and normalized such that their integral is unity. The inset shows an example of the re-scaling transformation with $\alpha=1.25$ in a $4.8 E_r$ deep lattice.}\label{self_similarity_graph}
\end{figure}

Another important conclusion from Eq. \ref{scaling} is that the distribution should exhibit a self-similar scaling with respect to $x t^{-1/\alpha}$, regardless of its exact shape. An example of this property is shown in the inset of figure \ref{self_similarity_graph}. When using the appropriate $\alpha$, all experimental data taken at different times collapse to the same curve when the x and y axes are re-scaled according to Eq. \ref{scaling}. This property can be used to extract the dynamical exponent; we employ an $L_1$-type measure of the self-similarity:
\begin{equation}\label{selfsimilarity_measure}
m(\alpha)=\sum_i\frac{\int_{-\infty}^{\infty} |\tilde{W}_i(x,\alpha)-\overline{W}(x,\alpha)|dx}{\int_{-\infty}^{\infty} \overline{W}(x,\alpha)dx} \ \ ,
\end{equation}
where $\tilde{W}_i(x,\alpha)=t_i^{1/\alpha}W(x t_i^{1/\alpha},t_i)$ are re-scaling of the series of distributions $W(x,t_i)$ measured at different times, and $\overline{W}(x,\alpha)$ is their average. In figure \ref{self_similarity_graph} we depict this measure as a function of $\alpha$ for three lattice depths. For each of these curves there is a single minimum whose position is shifted in accordance with the lattice depth. We interpret the minimum as the most probable value for the diffusion exponent. This value is plotted as a function of the lattice depth in figure \ref{diffusion_exponent}. The diffusion exponents obtained by the self-similarity method and by fitting the FWHM to a power-law agree to within the uncertainty.

\begin{figure}
    \begin{center}
    \includegraphics[width=7.5cm]{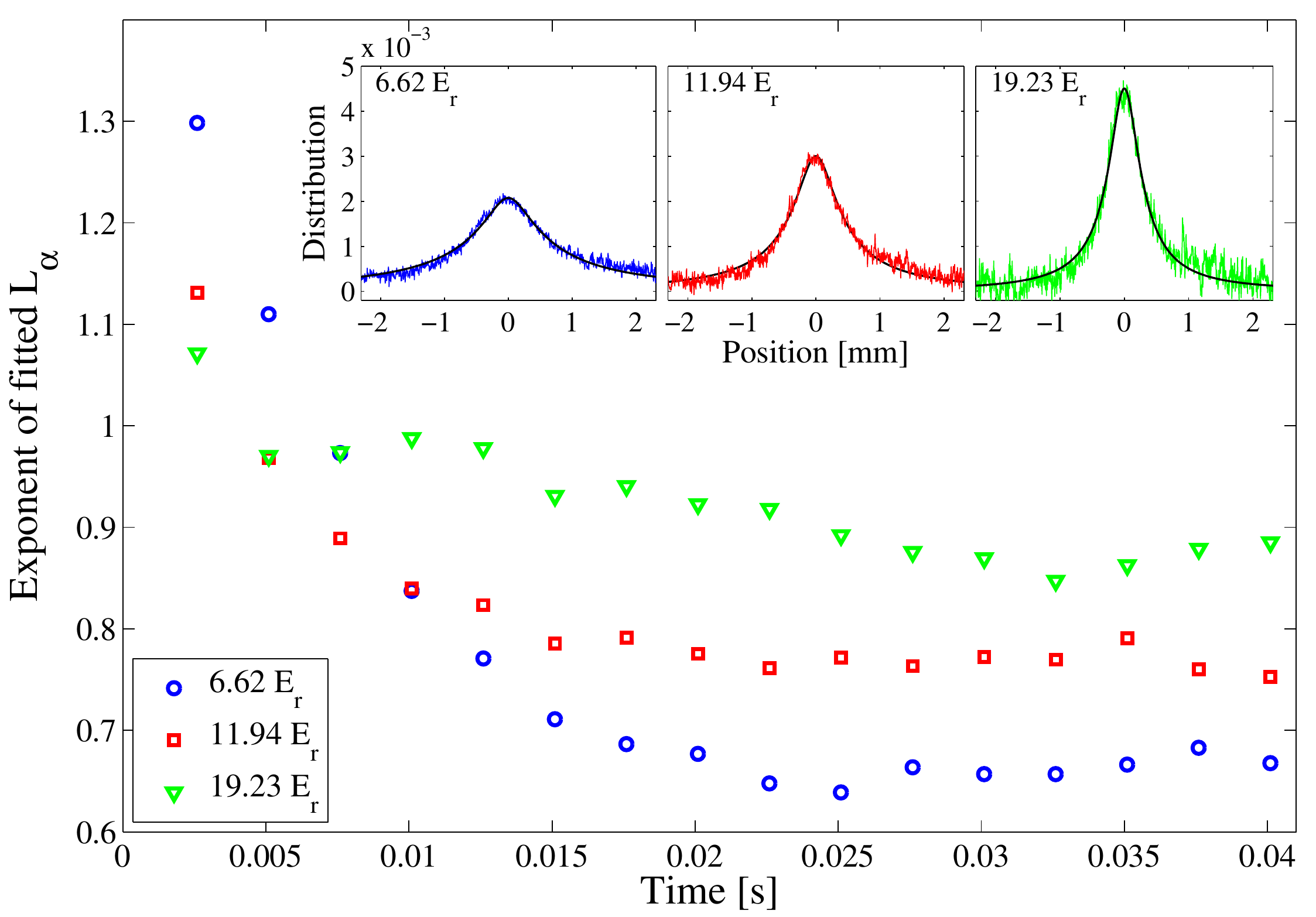}
    \end{center}\caption{The atomic distributions after different diffusion times are fitted with a L\`{e}vy $L_\alpha$ function. The resulting exponent $\alpha$ is plotted as a function of the diffusion time for three characteristic lattices. In the inset the distributions after 30ms of diffusion and their corresponding fits are   shown.}\label{fit_to_Levy}
\end{figure}

Up to this point we have analyzed the temporal behavior of the atomic distribution, and now we turn to study its shape. Motivated by the fact that for $\beta=1$ the solution for the kernel of Eq. \ref{FDE} is the L\'{e}vy stable law $L_\alpha$ \cite{Metzler20001,Mainardi2005}, we use the latter as a fitting function. In figure \ref{fit_to_Levy} we depict the shape exponent, $\alpha$, extracted from these fits as a function of the diffusion time, for 3 characteristic lattices. In all three lattices the shape exponent converge to an asymptotic value on a timescale of $10ms$. The inset shows the fits after $30ms$ of diffusion. The L\'{e}vy distributions fit the data very well with an average r-square of 0.96 and an uncertainty in the fitted $\alpha$ at a $95\%$ confidence level of $\sim3\%$. The shape exponent is depicted as triangles in figure \ref{diffusion_exponent}. At very shallow lattices it approaches 2; this is expected since for any finite observation time there are lattices which are too weak for the atoms to reach equilibrium. In deeper lattices, on the other hand, we consistently find that the shape exponent is significantly smaller than the dynamical exponents, and in particular smaller than 1. Similar results are obtained if the tail of the spatial distributions is fitted with a power-law instead of with a L\'{e}vy distribution.

In order to better understand why the shape exponent is different than the two dynamical exponents, we have written both classical and quantum simulations \cite{EPAPS_note}. Based on their results we attribute the discrepancy between the exponents to the combination of two factors: $\beta<1$ in the FDE and correlations between the particle's velocity and flight duration. The quantum simulation we have performed is a Monte Carlo Wave-Functions (MCWF) of atoms in one dimensional polarization lattice with angular momentum $J_g=1/2$ to $J_e=J_g+1=3/2$ transitions \cite{Molmer:93}. The atomic evolution in the light field is treated quantum mechanically and photon scattering is described by quantum jumps. In general, we find a good qualitative agreement between experiments and the simulation despite differences in the level structure and lattice detuning which we introduce to simplify the numerics \cite{EPAPS_note}. The temporal evolution of the simulated spatial distributions, as well as their shape and dependence on the lattice depth, exhibit the same properties that were discussed above. In particular, similarly to figure \ref{diffusion_exponent}, we find that the spatial L\'{e}vy exponent is consistently smaller than the dynamical exponent for deep lattices. The simulations also establish a clear correlation between the velocity and flight-duration of the atoms. In other words, if an atom acquires a large momentum, it is more likely not to be trapped in a single lattice site for longer times. We have also ran classical Monte-Carlo simulations where the velocities, flight-durations and dwelling times in the lattice are all drawn from Levy distributions. Though this simulation shows that a shape exponent smaller than 1 can be obtained by $\beta<1$ for shallow lattices, it is necessary to include the correlations between the velocity and flight duration in order to sustain this result for increasingly deeper lattices. Both these factors were indeed found in the MCWF simulations \cite{EPAPS_note}.

To summarize, we have presented measurements and analysis of spatial anomalous diffusion of ultra-cold atoms in a 1D polarization lattice. We find that a complete description of this process goes beyond the FDE and must include the effect of correlations between the motion variables. Future extensions of our work include the study of anomalous diffusion in the presence of an external force and diffusion in very shallow lattices where there is hope to observe super-ballistic diffusion \cite{PhysRevA.83.043821}. Also of great interest is to measure the time-evolution of the velocity distribution which is also expected to exhibit anomalous diffusion \cite{PhysRevLett.105.120602,PhysRevE.84.041111}.

We thank Rami Pugatch for stimulating discussions. This work was partially supported by MIDAS, MINERVA, ISF and DIP.


%

\end{document}